\documentclass[aps,prl,amssymb,showpacs,superscriptaddress,twocolumn]{revtex4}
\usepackage{graphicx}
\def\rv{{\bf r}}

\def\kv{{\bf k}}
\def\qv{{\bf q}}
\def\0v{{\bf 0}}
\def\Av{{\bf A}}

\def\Gv{{\bf G}}

\begin{document}

\title{Exact dynamical exchange-correlation kernel of a weakly inhomogeneous electron gas}

\author{V.~U.~Nazarov}

\affiliation{Research Center for Applied Sciences, Academia Sinica,
 Taipei 115, Taiwan}
\affiliation{Department of Physical Chemistry, Far-Eastern National Technical University, 10 Pushkinskaya Street, Vladivostok 690950, Russia}

\author{G.~Vignale}
\affiliation {Department of Physics and Astronomy, University of Missouri, Columbia, Missouri 65211, USA}

\author{Y.-C.~Chang}
\affiliation{Research Center for Applied Sciences, Academia Sinica,
 Taipei 115, Taiwan}

\begin{abstract}
The  dynamical exchange-correlation kernel $f_{xc}$ of a non-uniform electron gas is an essential input for the time-dependent density functional theory of electronic systems.  The long-wavelength behavior of this kernel is known to be of the form $f_{xc}= \alpha/q^2$ where $q$ is the wave vector and $\alpha$ is a frequency-dependent coefficient.  We show that in the limit of weak non-uniformity the coefficient $\alpha$  has a simple and exact expression in terms of the ground-state density and the frequency-dependent kernel of a {\it uniform} electron gas at the average density.  We present an approximate evaluation of this expression for Si  and discuss its implications for the theory of excitonic effects.
\end{abstract}

\pacs{31.15.ee}

\maketitle

\newcounter{Insulator}
Since its introduction in works of Runge, Gross, and Kohn \cite{Runge-84,Gross-85},
the time-dependent density-functional theory  (TDDFT) has evolved into a powerful tool of investigation
%in physics and chemistry, and
of systems ranging from isolated atoms to bulk solids.
In the important linear-response regime, the key quantity of TDDFT  is the dynamical
exchange-correlation (xc) kernel $f_{xc}$ defined as the functional derivative
\begin{eqnarray*}
f_{xc}[n_0({\bf r})]({\bf r},{\bf r}',\omega)=\left. \frac{\delta V_{xc}[n]({\bf r},\omega)}{\delta n({\bf r}',\omega)}\right|_{n=n_0({\bf r})}
\end{eqnarray*}
of the dynamical exchange and correlation potential $V_{xc}$ with respect to the dynamical electron density $n$,
taken at the ground-state value $n_0$ of the latter.
With this definition, the density-response
function $\chi$
can be represented  in operator notation as \cite{Gross-85}
\begin{eqnarray}
\chi(\rv,\rv',\omega)=\left\{[1-\chi_{KS}(C+f_{xc})]^{-1} \chi_{KS}\right\}(\rv,\rv',\omega)\,,
\label{oper}
\end{eqnarray}
where $\chi_{KS}$ is the Kohn-Sham (KS) density-response function of independent electrons, $C=e^2/|{\bf r}-{\bf r}'|$ is the Coulomb interaction, and $e$ is the absolute value of the electron charge.
While the density-response function of non-interacting electrons $\chi_{KS}$ can be straightforwardly calculated
in many cases of interest (e.g., for homogeneous electron gases in three and two dimensions it is given by the analytical Lindhard's \cite{Lindhard-54} and Stern's \cite{Stern-67} formulas, respectively),
the construction of $f_{xc}$, whose role is to account for dynamical many-body correlations, is not  straightforward.

As an instructive specific case, let us consider the excitonic effect \cite{Ashcroft&Mermin} in a semiconductor,
which would manifest itself as an enhancement of the imaginary part of $\chi$ for frequencies close to the fundamental absorption edge.   We neglect for a moment  local-field effects  and write down the diagonal elements of the density response in momentum space as
of Eq.~(\ref{oper})
\begin{eqnarray}
\chi(\qv,\qv,\omega)=\frac{\chi_{KS}(\qv,\qv,\omega)}{1-\chi_{KS}(\qv,\qv,\omega)[\frac{4\pi e^2}{q^2}+f_{xc}(\qv,\qv,\omega)]}\,
\label{opers}
\end{eqnarray}
where $4 \pi e^2/q^2$ is the Fourier transform of the Coulomb interaction.
On the one hand,  the excitonic enhancement of $\chi$ is a many-body effect and, therefore,
it needs a nonzero $f_{xc}$  to be accounted for within TDDFT.
On the other hand, because of the divergent Coulomb part $4\pi e^2/q^2$ in Eq.~(\ref{opers}), any $f_{xc}(q,q,\omega)$ that remained finite at $q=0$
would  give no contribution in the long-wave limit $q\rightarrow 0$. This simple observation  shows that in order  to include the exciton, $f_{xc}(\qv,\qv,\omega)$
must be divergent in the long-wave limit at least as strongly as the Coulomb term. And indeed, when the $q^{-2}$ divergence has been introduced  empirically in  papers dealing with the optical absorption spectrum of semiconductors~\cite{Reining-02,Del_Sole-03,Botti-04}, it has yielded a good TDDFT description of the excitonic effect.

Clearly it would be highly desirable to have a first-principle theory of the small-$\qv$ behavior of the xc kernel, rather than relying on empirical parametrizations.  In this Letter we take a step in this direction.  We first show that the asymptotic relation
\begin{eqnarray}
\lim_{\qv\rightarrow 0} f_{xc}(\qv,\qv,\omega)=\frac{e^2 \alpha(\omega)}{q^2}
\label{as}
\end{eqnarray}
(we introduce the $e^2$ so that $\alpha$ is dimensionless) is a rigorous consequence of exact sum rules for the current density response function.  Then, in the limit of weak non-uniformity we obtain a simple and exact expression for $\alpha(\omega)$ in terms of the ground-state density and the dynamical xc kernel of a homogeneous electron gas at the average density\footnote{The existence of a divergence in the {\it off-diagonal} components $f_{xc}(\qv,\kv+\qv,\omega)$ for $\qv\to 0$ and $\kv$ finite was first pointed out in Ref.~\onlinecite{Vignale-96}.}.

We start by noting that the local density approximation (LDA) to $f_{xc}$ is unable  to produce the divergence.
Indeed, within LDA \cite{Gross-85}
\begin{eqnarray}
f_{xc}({\bf r},{\bf r}',\omega)= f_{xc}^h[n({\bf r}),\omega] \, \delta({\bf r}-{\bf r}'),
\label{LDA}
\end{eqnarray}
where $f_{xc}^h(n,\omega)$ is the long-wave limit of the xc kernel of  the homogeneous electron gas of  density $n$.
The latter is known
\cite{Gross-85,Nifosi-98,Conti-99,Qian-02} to be finite, no divergence arising, therefore, in the Fourier transform of Eq.~(\ref{LDA}).

To obtain an accurate non-local $f_{xc}$, we  resort to the recently proposed general method \cite{Nazarov-07}
derived from the time-dependent {\em current}-density functional theory (TDCDFT) \cite{Vignale-96}.
This method is based on the exact relation that holds between the scalar density-response function $\chi$ (density response to a scalar potential)
and the tensor current-density-response function $\hat{\chi}$ (current-density response to a vector potential):
\begin{equation}\label{GaugeRelation}
\chi(\qv,\qv',\omega)=\frac{c}{e\omega^2}\qv \cdot \hat \chi(\qv,\qv',\omega) \cdot \qv'.
\label{rel}
\end{equation}
Both response functions are expressed in terms of the corresponding Kohn-Sham response functions and xc kernels in the following manner:
\begin{equation} \label{Chifxc1}
\chi^{-1}(\qv,\qv',\omega)=\chi_{KS}^{-1}(\qv,\qv',\omega)-f_{xc}(\qv,\qv',\omega)-\frac{4\pi e^2}{q^2}\delta_{\qv\qv'}
\end{equation}
and
\begin{equation} \label{Chifxc2}
\hat\chi^{-1}(\qv,\qv',\omega)=\hat\chi_{KS}^{-1}(\qv,\qv',\omega)-\hat f_{xc}(\qv,\qv',\omega)-\frac{4\pi e c}{\omega^2}\hat L_{\qv}\delta_{\qv\qv'}\,,
\end{equation}
where $\hat L_{\qv,ij} \equiv q_iq_j/q^2$, $i$ and $j$ are cartesian indices.

Equations~(\ref{GaugeRelation}-\ref{Chifxc2}) establish a connection between $f_{xc}$ and its tensor counterpart $\hat f_{xc}$.   The usefulness of this connection stems from the fact that the tensor quantities $\hat \chi_{KS}$ and $\hat f_{xc}$ satisfy a broader set of exact sum rules  than the corresponding scalar quantities. These sum rules were derived in Ref. \cite{Vignale-B}.  Specializing to the case of a periodic systems, the two most important sum rules for our purposes are
\begin{eqnarray}\label{SumRule1}
&&\hat{\chi}_{KS,ij}({\bf G},0,\omega)    =  \frac{e}{m c} n_0({\bf G}) \delta_{ij}\cr
&&- \frac{1}{m \omega^2}   \sum\limits_{{\bf G}',k}  \hat{\chi}_{KS,i,k}({\bf G},{\bf G}',\omega) G'_k G'_j V_{KS}({\bf G}')\,,
\end{eqnarray}
and
\begin{eqnarray}\label{SumRule2}
\sum\limits_{ {\bf G}'}\hat{f}_{xc,ij}({\bf G},{\bf G}',\omega) n_0({\bf G}') = \frac{c}{e \omega^2} \, G_i G_j V_{xc}({\bf G})\,,
\end{eqnarray}
where $\Gv$ are reciprocal lattice vectors.
These sum rules connect three different types of components of, say, $\chi_{KS}(\Gv,\Gv',\omega)$: the  $({\bf 0},{\bf 0})$ component,  the  $({\bf 0},{\Gv\neq{\bf 0}})$ and  $({\Gv\neq{\bf 0}},{\bf 0})$ components, and the  $({\Gv\neq{\bf 0}},{\Gv'\neq{\bf 0}})$ components.

Let us further restrict our attention to the case of a weakly inhomogeneous system: $|n_0(\Gv)|\ll n_0({\bf 0}) \equiv \bar n_0$, and $|V_{KS}(\Gv)| \ll \hbar^2G^2/2m$  for $\Gv \neq {\bf 0}$.   Then it is easily shown that the  homogenous electron gas approximation for the $({\Gv\neq{\bf 0}},{\Gv'\neq{\bf 0}})$ components  completely determines the $({\bf 0},{\Gv\neq{\bf 0}})$ and  $({\Gv\neq{\bf 0}},{\bf 0})$ components to {\it first order} in $n_0(\Gv \ne {\bf 0})$, which in turn completely determines the $({\bf 0},{\bf 0})$ component  to {\it second order} in $n_0(\Gv \ne {\bf 0})$.  Thus, for $\hat{\chi}_{KS}$ we obtain
\begin{widetext}
\begin{eqnarray}
\hat{\chi}_{KS,ij}({\bf G}\ne \0v,{\bf G}'\ne \0v,\omega)&=&
\left[ \frac{e \omega^2}{c\, G^2} L_{{\bf G},ij} \, \chi^{hL}_{KS}(G,\omega)+T_{{\bf G},ij} \, \chi^{hT}_{KS}(G,\omega)
\right] \delta_{{\bf G G}'}\,,\cr\cr
\hat{\chi}_{KS,ij}({\bf G} \ne \0v,\0v,\omega)&=&\hat{\chi}_{KS,ij}(\0v,-{\bf G},\omega)=
\frac{e}{m c} \left[n_0({\bf G}) \delta_{ij}
- L_{{\bf G},ij} \chi^{hL}_{KS}(G,\omega) V_{KS}({\bf G})\right]\,,\cr\cr
\hat{\chi}_{KS,ij}(\0v,\0v,\omega)&=& \frac{e \bar n_0}{m c } \delta_{ij} + \frac{e}{m^2 \omega^2 c} \sum\limits_{{\bf G}\ne \0v} G^2 L_{{\bf G},ij} |V_{KS}({\bf G})|^2
\left[\chi^{hL}_{KS}(G,\omega) -\chi^{hL}_{KS}(G,0)\right]\,.
\label{chiKS}
\end{eqnarray}
to the zero-th, first, and second order in $V_{KS}({\bf G})$, respectively. Here
$\chi^{hL}_{KS}$ and$\chi^{hT}_{KS}$ are, respectively, the longitudinal and transverse KS density-response functions of the homogeneous electron gas of density $\bar n_0$, and $T_{{\bf G},ij}=\delta_{ij} -L_{{\bf G},ij}$.
Similarly, for $\hat f_{xc}$ we have
\begin{eqnarray}
&&\hat{f}_{xc,ij}({\bf G}\ne 0,{\bf G}' \ne 0,\omega)= \frac{c}{e \omega^2 } \, G^2  \! \left[ f_{xc}^{hL}(G,\omega)  L_{{\bf G},ij} \! + \! f_{xc}^{hT}(G,\omega) T_{{\bf G},ij}   \right] \delta_{{\bf G G}'}\,,
%\label{fxct3}
\nonumber
\\
&&\hat{f}_{xc,ij}({\bf G}\ne \0v,\0v,\omega) \! = \! \hat{f}_{xc,ji}(\0v,-{\bf G},\omega) \! = \!
-\frac{c \, G^2}{e \omega^2 \bar n_0}  n_0({\bf G}) \! \left\{ [f_{xc}^{hL}(G,\omega) \! - \! f_{xc}^{hL}(G,0)] L_{{\bf G},ij} \! + \! f_{xc}^{hT}(G,\omega) T_{{\bf G},ij} \!  \right\}\,,
%\label{fxct2}
\nonumber
\\
&&\hat{f}_{xc,ij}(\0v,\0v,\omega)=
\frac{c}{e \omega^2 \bar n_0^2} \sum\limits_{ {\bf G}\ne \0v} G^2 |n_0({\bf G})|^2
\left\{ \left[f_{xc}^{hL}(G,\omega) -f_{xc}^{hL}(G,0) \right]L_{{\bf G},ij} + f_{xc}^{hT}(G,\omega) T_{{\bf G},ij} \right\},
%\label{fxct1}
\label{FXC}
\end{eqnarray}
\end{widetext}
where
$f_{xc}^{hL}$ and $f_{xc}^{hT}$  are the longitudinal and transverse, respectively, xc kernels of the homogeneous electron gas of  density $\bar n_0$.
\begin{figure}[h]
\includegraphics[width=0.4\textwidth,height=0.3\textwidth]{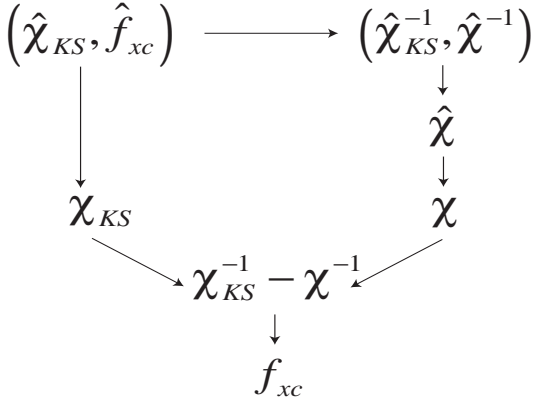}
\caption{\label{Flowchart}Scheme of the procedure for calculating the xc kernel $f_{xc}$ starting from the expressions~(\ref{chiKS}) and (\ref{FXC}) for $\hat \chi_{KS}$ and $\hat f_{xc}$, respectively.
}
\end{figure}

The following steps, which involve repeated inversions of infinite matrices,
rely on the mathematical fact that to find the $({\bf 0},{\bf 0})$, $({\bf 0},{\bf G}\ne {\bf 0})/({\bf G}\ne {\bf 0},{\bf 0})$, and $({\bf G}\ne {\bf 0},{\bf G}'\ne {\bf 0})$
elements of the inverse matrix to the second, first, and zeroth order in the inhomogeneity, respectively,
it is sufficient to know the corresponding elements of the original matrix
to the same accuracy, and then the inversion can be performed in a closed form \cite{Sturm-82}.

The complete procedure is schematically illustrated in Fig.~(\ref{Flowchart}). Starting from Eqs.~(\ref{chiKS}) and~(\ref{FXC}) for $\hat \chi_{KS}$ and $\hat f_{xc}$ we (i)  Invert Eqs.~(\ref{chiKS}) to get $\hat{\chi}_{KS}^{-1}$; (ii) Combine  $\hat{\chi}_{KS}^{-1}$ and $\hat f_{xc}$ to get
$\hat{\chi}^{-1}$ by virtue of Eq.~(\ref{Chifxc2}); (iii) Invert $\hat{\chi}^{-1}$ to get $\hat \chi$ ; (iv) Use Eq.~(\ref{GaugeRelation}) and its KS analogue to find the scalar response function $\chi$ from $\hat{\chi}$
and $\chi_{KS}$ from $\hat{\chi}_{KS}$;
(v) Invert $\chi$ and $\chi_{KS}$  to get $\chi^{-1}$ and $\chi_{KS}^{-1}$;
and, (vi) Apply Eq.~(\ref{Chifxc2}) to find $f_{xc}$.
The final result of this procedure is
\begin{widetext}
\begin{eqnarray}
&& \lim_{q \rightarrow 0} f_{xc}({\bf G}\ne \0v,{\bf G}' \ne \0v,\omega) = f^h_{xc,L}(G,\omega) \delta_{{\bf G G}'},
\label{fxc3}\\
&& \lim_{q \rightarrow 0} f_{xc}({\bf G} \ne \0v,{\bf q},\omega) = f_{xc}(-{\bf q},-{\bf G},\omega)= - \frac{({\bf G}\cdot {\hat{\bf q}})}{\bar n_0 q}
\left[f_{xc}^{hL}(G,\omega) -f_{xc}^{hL}(G,0) \right] n_0({\bf G}),
\label{fxc2} \\
&& \lim_{q \rightarrow 0} f_{xc}({\bf q},{\bf q},\omega) =
\frac{1}{\bar n_0^2 q^2}
\sum\limits_{{\bf G}\ne \0v} ({\bf G}\cdot {\hat{\bf q}})^2 \left[f_{xc}^{hL}(G,\omega) -f_{xc}^{hL}(G,0) \right] |n_0({\bf G})|^2,
\label{fxc}
\end{eqnarray}
\end{widetext}
where $\hat{\bf q}$ is the unit vector parallel to ${\bf q}$. It should be noted at this point that the above expression for the scalar kernel $f_{xc}({\bf q},{\bf q},\omega)$  differs from what one would get by simply taking the longitudinal component of $\hat f_{xc,ij}({\bf q},{\bf q},\omega)$, i.e.  $f_{xc}({\bf q},{\bf q},\omega) \neq \frac{e\omega^2}{c q^2}\sum_{i,j} \hat q_i \hat{f}_{xc,ij}({\bf q},{\bf q},\omega) \hat q_j$.  The implication is that the scalar xc potential ($V_{xc}$) of time-dependent DFT is {\it not} equivalent to the longitudinal component of  the vector potential ($\Av_{xc}$) of  time-dependent CDFT: rather, it should be constructed through the careful inversion procedure described above.  A recent interesting attempt to construct  $V_{xc}$ from $\Av_{xc}$~\cite{Maitra-07} should be re-examined in the light of this result.

From the result of the step (iv) for $\chi$ and making use of the relation
\begin{eqnarray}
\frac{1}{\epsilon_M(\omega)}= 1+\lim_{q\rightarrow 0} \frac{4\pi e^2}{q^2} \chi({\bf q},{\bf q},\omega),
\label{eMM}
\end{eqnarray}
we obtain a formula for the macroscopic dielectric function of a crystal
\begin{eqnarray}
&& \! \! \! \! \! \! \! \! \! \! \! \epsilon_M(\omega)  =  1  -
\frac{4\pi e^2 \bar n_0}{ m \omega^2 }     -
 \frac{ e^2} {  m^2 \omega^4  } \cr\cr
 && \! \! \!  \! \! \! \! \! \! \! \! \! \! \! \! \times \sum\limits_{{\bf G}\ne 0}  |V_{0}({\bf G})|^2 G^2
(\hat{{\bf q}}  \cdot  {\bf G})^2
 \left[ \frac{1}{\epsilon^{hL}(G,\omega)}- \frac{1}{\epsilon^{hL}(G,0)}\right] \! \! ,
\label{eM}
\end{eqnarray}
where $V_0$ is the bare crystalline potential and
\begin{eqnarray*}
\epsilon^{hL}(q,\omega)=1-\frac{4\pi e^2}{q^2} \frac{\chi^{hL}_{KS}(q,\omega)}{1-\chi^{hL}_{KS}(q,\omega) f^h_{xc,L}(q,\omega)}
\end{eqnarray*}
is the longitudinal dielectric function of the homogeneous  electron liquid.
Equation (\ref{eM}) is in agreement with the Hopfield's formula for optical conductivity \cite{Hopfield-65},
while in the RPA [$f^h_{xc,L}(G,\omega)=0$] it coincides with the corresponding result of Ref.~\onlinecite{Sturm-82}.

Equations~(\ref{fxc3}-\ref{fxc}) are the main result of this paper.
They  replace the grossly inaccurate LDA formula
\begin{eqnarray}
\lim_{q \rightarrow 0} f_{xc}({\bf G}+{\bf q},{\bf G}'+{\bf q},\omega) = f^h_{xc,L}(G,\omega) \delta_{{\bf G G}'},
\label{fxc33}
\end{eqnarray}
which does not contain the singularity in $q$.
Identifying the ($\0v,\0v$) element
of the microscopic matrix of the xc kernel in Eqs.~(\ref{fxc}) as the averaged $f_{xc}$, we see that
$f_{xc}$ diverges for $q\rightarrow 0$ as described by Eq.~(\ref{as}), wherein
$\alpha(\omega)$ is given by
\begin{eqnarray}
\alpha(\omega) \! \! = \! \!
\! \! \sum\limits_{{\bf G}\ne 0} \! \! \! \frac{({\bf G} \! \cdot \! \hat{{\bf q}})^2 }{ \bar n_0^2} \! \left[f_{xc}^{hL}(G,\omega) \! - \! \! f_{xc}^{hL}(G,0) \! \right] \! |n_0({\bf G})|^2.
\label{alpha}
\end{eqnarray}
Notice that $\alpha(\omega)=0$ in the uniform limit and  $\alpha(0)=0$ up to second order in $n_0(\Gv \ne {\bf 0})$
\footnote{This is not to say that $f_{xc}(\qv,\qv,0)$ is always free of the $q^{-2}$ singularity, but it means that such a singularity, if present,  cannot be reached by a perturbative expansion about the homogeneous ground-state.  Indeed, there is evidence that  the $f_{xc}(\qv,\qv,0)$ of band insulators has a $q^{-2}$ singularity, which  is missed in the present approach.}.
\setcounter{Insulator}{\value{footnote}}

In order to calculate $\alpha(\omega)$ from Eq.~(\ref{alpha}) we need the Fourier amplitudes of the ground-state electron density and the wave vector and frequency-dependent $f_{xc}^{hL}$ of the homogeneous electron gas, evaluated at reciprocal lattice vectors.  The first ingredient is straightforwardly obtained from standard electronic structure calculations.  Unfortunately,  the same cannot be said of the second ingredient $f_{xc}^{hL}(q,\omega)$, for which we do not have reliable expressions.  The best that can be done, at this time, is either to disregard the wave vector dependence, or to make use of the interpolation formula proposed in Ref.~\onlinecite{Dabrowski-86}, which however fails to reproduce, at small $q$, what is presently believed to be the qualitatively correct form of the frequency dependence.  In spite of these difficulties, it must be emphasized that the calculation of $f_{xc}^{hL}(q,\omega)$ is still a much simpler problem than the calculation of the dynamical xc kernel of the non-uniform system.  Thus, our Eq.~(\ref{alpha}) does not simply express an unknown quantity in terms of another unknown quantity, but actually opens the way to systematic calculations of $\alpha$  based on the many-body theory of the homogeneous electron gas.   Further, Eqs. (\ref{fxc3})-(\ref{fxc}) for $f_{xc}$ offer a promising alternative to the widespread practice of treating the dynamical exchange and correlations effects in the LDA.

\begin{figure}[h]
\includegraphics[width=0.475\textwidth,height=0.35\textwidth]{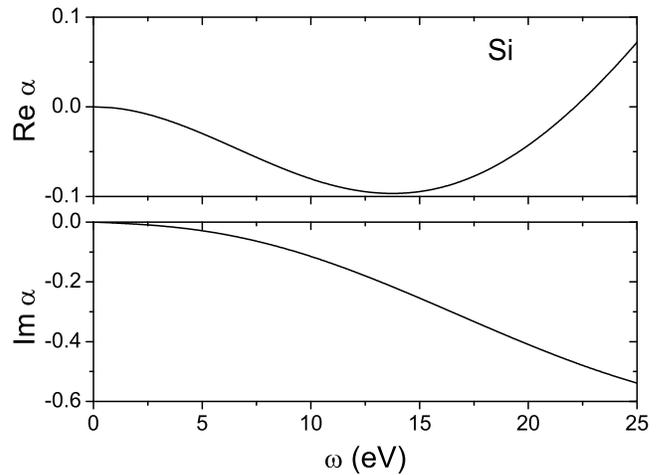}
\caption{\label{Fig}
The frequency dependence of the real (upper panel) and image (lower panel) parts of the $\alpha$ coefficient in Eq.~(\ref{as})
for silicon calculated by Eq.~(\ref{alpha}).
}
\end{figure}

In Fig.~\ref{Fig} we plot $\alpha(\omega)$ from Eq.~(\ref{alpha}) vs frequency for crystalline silicon.
The Fourier coefficients of the electron density were calculated with the code FHI98MD~\cite{Bockstedte-97}, and we approximated  $f_{xc}^{hL}(q,\omega) \simeq f_{xc}^{hL}(0,\omega)$,
taking the latter from Ref.~\onlinecite{Qian-02}.
In the range 0-22 eV, the real part of $\alpha(\omega)$ is negative, changing sign for positive above 22 eV.
It reaches its minimum of $\alpha \approx$  -0.1 at $\omega\approx$ 14 eV.
In the range 3-5 eV of the main absorption in silicon, ${\rm Re} \, \alpha$ changes from -0.01 to -0.03,
which is an order of magnitude smaller than the empirical value of $\alpha \approx$ -0.2
found as the best fit to the experimental spectrum in Ref.~\onlinecite{Reining-02}. This large difference may simply indicate that the nearly free electron model, while being adequate for simple metals and even for semiconductors
in the high-frequency regime \cite{Sturm-82}, is not sufficiently accurate for semiconductors at frequency lower than or comparable to the band gap (see also
footnote~[\arabic{Insulator}]).  Another probable source of discrepancy is that
our approach is a pure TDDFT, whereas the value of $\alpha \approx -0.2$ was obtained in Refs.~\onlinecite{Reining-02} and~\onlinecite{Del_Sole-03} with the use of self-energies incorporated in the Green's function via the $GW$ approximation.

In conclusion,
within the nearly free electron approximation, we have constructed the otherwise exact
exchange-correlation kernel  for time-dependent density-functional theory.
This kernel is nonlocal in space, exhibiting the $q^{-2}$ singularity in the reciprocal space.
The strength of this singularity, which is frequency-dependent,
has been directly related to the magnitude of the non-uniformity of the density of valence electrons,
and this singularity disappears in the limiting case of the homogeneous electron liquid.
We are proposing an improvement over the conventional LDA scheme of including the dynamical exchange and correlation effects
into {\em ab initio} calculations of the linear response of crystalline solids
which consistently accounts for the long-wave divergence in the exchange-correlation kernel.

GV acknowledges support from DOE Award No. DE-FG02-05ER46203.

\end{document}